\documentclass[aps,pra,twocolumn,showpacs,amssymb,amsmath,amsfonts,superscriptaddress,floatfix,nofootinbib,groupedaddress]{revtex4-1}
\usepackage{graphicx}
\usepackage{mathtools}
\usepackage{perpage} 
\usepackage{amsmath,amssymb}
\MakePerPage{footnote}
\usepackage{color}

\begin{document}

%\preprint{}

\title{Optimal exploitation of the resource in remote state preparation}


\author{Morteza Nikaeen$^{1}$, Mehdi Ramezani$^{1,2}$ and Alireza Bahrampour$^{1}$}

\affiliation{$^{1}$Department of Physics, Sharif University of Technology, Tehran 14588, Iran\\$^{2}$School of Physics, Institute for Research in Fundamental Sciences (IPM), Tehran 19395, Iran}

\date{\today}

\begin{abstract}
Transmission efficiency (TE) of remote state preparation (RSP) with a shared quantum state and one bit of classical communication is considered.  Following [B. Dakić et al., Nat. Phys. 8, 666 (2012)], the encoding and decoding strategies of the protocol are restricted to the physically relevant classes of projective measurements and unitary operators, respectively. It is shown that contrary to the previous arguments, the quadratic fidelity as well as the linear fidelity could be a valid figure of merit to quantify the TE of RSP. Then, the TE of the protocol in terms of both linear and quadratic fidelities is evaluated in a fully optimized scenario which includes the maximization over the encoding parameters as well as a meaningful maximization over the decoding parameters. The results show that in this scenario, the TE scales with the sum of the two largest eigenvalues of the squared correlation matrix of the resource state that is zero only for product states. This approach successfully quantifies the performance of the protocol in terms of the resource state parameters and provides a means to compare the usefulness of any two resource states for RSP. 
\end{abstract}

%\keywords{}

\maketitle


%%%%%%%%%%%%%%%%%%%%%%%%%%%%%%%%%%%%%%%%%%%%%%%%%%%%%%%%
%%%%%%%%%%%%%%%%%%%%%%%%%%%%%%%%%%%%%%%%%%%%%%%%%%%%%%%%

\section{Introduction}\label{Introduction}
Quantum communication theory is based on the fact that quantum correlations can be exploited for performing communication tasks that are classically impossible or less efficient. However, when considering composite systems and analyzing the nature of correlations between their subsystems, recognizing that when they behave in a quantum way, rather than as an effectively classical one, lacks a universal answer. In fact, after recognizing every genuine quantumness in composite quantum systems, one can quantify the quantumness of the correlations using various distance measures. This is why there exists a zoo of quantum correlations measures \cite{adesso2016measures}.\

Among all proposed measures of quantum correlations, one is more attractive, that can be interpreted operationally. For the most famous quantum correlation measures, i.e. quantum discord \cite{henderson2001classical}, several operational interpretations have been presented through some quantum communication tasks such as state merging \cite{madhok2011interpreting}, dense coding \cite{gu2012observing}, quantum locking \cite{nikaeen2018quantum} and remote state preparation \cite{dakic2012quantum}.\

Remote state preparation (RSP) is one of the significant building blocks in quantum communication theory \cite{pati2000minimum}, \cite{bennett2001remote}. In this task, a qubit chosen from a certain great circle on a Bloch sphere can be remotely prepared with one classical bit from Alice to Bob if they share one ebit of entanglement. To recognize the optimal resource of RSP, the performance of the protocol has been evaluated for general shared resource states in terms of some quantum correlations measures.\

In the first approach \cite{dakic2012quantum}, a quadratic fidelity was adopted to quantify the transmission efficiency (TE) of RSP. Then, the TE of the protocol in the worst case scenario, i.e. when the worst great circle is chosen for sending the quantum states, was linked to the geometric discord of the resource state \cite{dakic2010necessary}. This result was supported experimentally, where certain separable states were shown to perform RSP better than some entangled states. Besides  other criticisms \cite{tufarelli2012quantum}, \cite{giorgi2013quantum}, this challenging result was mainly criticized in \cite{horodecki2014can}, where it was argued that using the quadratic fidelity as quantifier of the TE can be misleading, since regardless of the resources used in the protocol, it can be surpassed simply by employing a trivial guessing protocol.\

Recently, this approach has been followed again in \cite{kanjilal2018remote}, where it was shown that discord cannot quantify the
performance of the protocol for every great circle and fails as the quantifier of RSP performance to compare between usefulness of two resource states for RSP. To improve the approach, it was suggested that for a given resource state, instead of considering the worst case scenario, the average performance of the protocol should be considered. This improvement showed that for Bell-diagonal states, the TE of the protocol in terms of average optimal quadratic fidelity is in direct correspondence with an appropriate measure of quantum correlations known as SCMUB \footnote{Simultaneous Correlations in Mutually Unbiased Bases} \cite{guo2014quantum}, \cite{wu2014reveal}, which is zero only for product states. Further, given any two nonzero-discord states as a resource for RSP, it was shown that the SCMUB can successfully determine which particular resource state can be more useful than the other.\

Approaches considered in \cite{dakic2012quantum} and \cite{kanjilal2018remote}  use the quadratic fidelity to quantify the TE of RSP and utilize non-optimial unitary operators as decoding strategy. Furthermore, as will be discussed in this paper, the approach employed in \cite{kanjilal2018remote}, although mathematically sound, is not justified in the perspective of the resource theory.\\
To seek the actual relation between the TE of RSP and parameters of the resource, and consequently recognizing the ultimate resource of the task, in addition to the optimization over the encoding strategies, the optimization over the decoding strategies is also necessary. One can restrict the encoding and decoding operators to some special CPTP maps and seek the optimal resource of RSP in the corresponding class of the encoding-decoding maps. Since unitary operators are the most common operations implemented in the laboratories, and also RSP in its ideal form employs projective measurements and unitary operators, following \cite{dakic2012quantum} and \cite{kanjilal2018remote} we restrict the encoding and decoding operators to the class of projective measurements and unitary operators, respectively. In this context, first it is shown that contrary to the argument given in \cite{horodecki2014can}, the quadratic fidelity as well as the linear fidelity could be a valid figure of merit to quantify the TE of RSP. Then, the TE of the protocol in terms of both linear and quadratic fidelities is evaluated in a fully optimized scenario which includes maximization over the encoding parameters as well as a meaningful maximization over the decoding parameters. In this scenario, the TE scales with the sum of two largest eigenvalues of the squared correlation matrix of the resource state that is zero only for product states. Interestingly, in this approach all great circles exploit the resource state equally. Hence, this approach successfully quantifies the performance of the protocol for every great circle and provides a means to compare the usefulness of two resource states for RSP. Finally, the results are interpreted and compared with the results of the previous approaches. 
Our solutions are based on the combination of analytical and numerical methods that are detailed in Appendix.

%%%%%%%%%%%%%%%%%%%%%%%%%%%%%%%%%%%%%%%%%%%%%%%%%%%%%%%%
%%%%%%%%%%%%%%%%%%%%%%%%%%%%%%%%%%%%%%%%%%%%%%%%%%%%%%%%

\section{Standard remote state preparation}\label{RSP}
Remote state preparation \cite{pati2000minimum} in its standard form involves two parties, Alice (A) as the sender and Bob (B) as the  receiver of quantum information. Alice and Bob have access to the shared maximally entangled state ${\left| \psi  \right\rangle _{AB}} = \frac{{\left| {01} \right\rangle  - \left| {10} \right\rangle }}{{\sqrt 2 }}$, and a classical communication channel. In this task Alice aims to remotely prepare Bob's qubit system in the signal quantum state
\begin{equation}\label{QSS}
\left| {{\psi _s}} \right\rangle  = \frac{{\left| 0 \right\rangle  + {e^{i\varphi }}\left| 1 \right\rangle }}{{\sqrt 2 }},
\end{equation}
which is known to Alice and is unknown to Bob and lies on the equator of the Bloch sphere. The equator is characterized by direction $\hat z$ that is the positive eigen-state of Pauli operator $\sigma _{z}$. The singlet shared state can be equivalently written (up to a global
phase)  in the $\{ \left| {{\psi _s}} \right\rangle ,{\left| {{\psi _s}} \right\rangle ^ \bot }\}$ basis as
\begin{equation}\label{CRSP}
{\left| \psi  \right\rangle _{AB}} = \frac{{\left| {{\psi _s}} \right\rangle {{\left| {{\psi _s}} \right\rangle }^ \bot } - {{\left| {{\psi _s}} \right\rangle }^ \bot }\left| {{\psi _s}} \right\rangle }}{{\sqrt 2 }},
\end{equation}
where ${\left| {{\psi _s}} \right\rangle ^ \bot }$ is the state orthogonal to the signal quantum state. To remotely prepare Bob's state, Alice applies a von Neumann measurement in the basis $\{ \left| {{\psi _s}} \right\rangle ,{\left| {{\psi _s}} \right\rangle ^ \bot }\}$ on her part of the singlet. Depending on her outcome, Bob's system is found in one of the states $\left| {{\psi _s}} \right\rangle$ or ${\left| {{\psi _s}} \right\rangle ^ \bot }$. By sending the outcome of her measurement to Bob, which implies sending one classical bit, he either finds his system in the desired state $\left| {{\psi _s}} \right\rangle$, or can correct his state ${\left| {{\psi _s}} \right\rangle ^ \bot }$ by applying the Pauli operator ${\sigma _z}$.\

RSP in its standard form is a deterministic task thanks to the maximally entangled shared state, i.e. Alice could remotely prepare the signal state with certainty. For general shared states, the most Alice can do is to adjust her measurement such that Bob's final state becomes as close as possible to the signal state. In what follows the TE of RSP for the general shared resource states will be considered. 

%%%%%%%%%%%%%%%%%%%%%%%%%%%%%%%%%%%%%%%%%%%%%%%%%%%%%%%%
%%%%%%%%%%%%%%%%%%%%%%%%%%%%%%%%%%%%%%%%%%%%%%%%%%%%%%%%

\section{Remote state preparation in the presence of noise and the efficiency of the protocol}\label{TE}
In this section our purpose is to characterize the TE of RSP for general resource states, including entangled as well as separable states, in terms of parameters of the resource state. In this regard, after presenting a systematic model of RSP, an overview on the previous approaches is given. These approaches use the quadratic fidelity as quantifier of the TE of RSP. The justification of using the quadratic fidelity is discussed. Then, the TE of RSP in terms of the linear and quadratic fidelities is optimized over the whole encoding-decoding parameters to get a relation between the optimized fidelities and parameters of the resource state. Finally, the decoding strategy employed in the previous approaches is also analyzed.

%%%%%%%%%%%%%%%%%%%%%%%%%%%%%%%%%%%%%%%%%%%%%%%%%%%%%%%%

\subsection{Model description}
\begin{figure}[htp]
	\centering
	\includegraphics[scale=0.5]{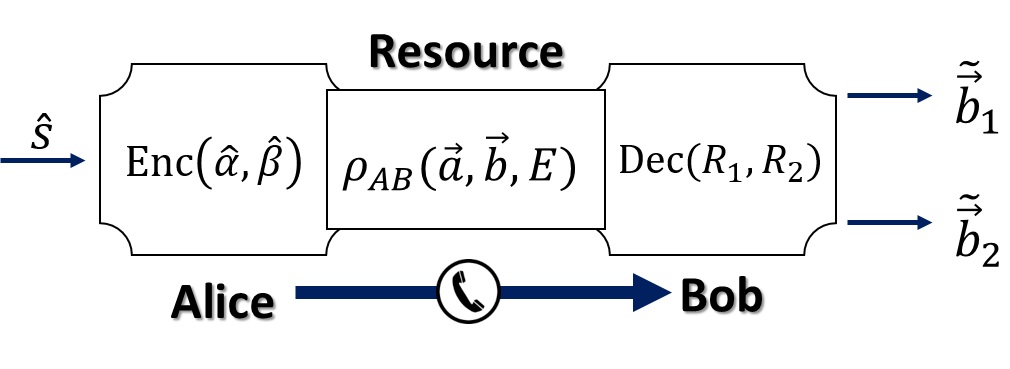}
	\caption{The schematic model of RSP. The protocol is modeled as an encoding-decoding mechanism that exploits the shared resource state and classical communication channel to perform the RSP task. The vectors $\hat s$ and ${{\tilde {\vec b}}_{1,2}}$ are Bloch vectors of input signal state and probabilistic output states, respectively. \label{Model}}
\end{figure}

To unify the previous studies with the current one, the RSP protocol is modeled as an encoding-decoding mechanism that exploits the shared resource state and one classical bit to perform the RSP task as depicted in Fig. \ref{Model}.\

In RSP protocol, Alice aims to transmit the signal state $\left| {{\psi _s}} \right\rangle$, that in the Bloch sphere representation, corresponds to a unit vector ${\hat s}$. According to the protocol, the signal state is selected from a certain great circle characterized by its normal vector ${\hat \beta }$ (in our notation $\hat s \in \mathrm{GC}(\hat \beta )$). The parameterization of $\hat s \in \mathrm{GC}(\hat \beta )$ in the spherical coordinate system is given in  Appendix \ref{parameterizationapp}.\

To perform the RSP task, Alice and Bob exploit shared resource state ${\rho _{AB}}$ as a passive quantum channel, where state ${\rho _{AB}}$ , in its general form can be written in two-qubit Bloch vector representation as
\begin{equation}\label{rho}
{\rho _{AB}} = \frac{1}{4}[I \otimes I + \vec a.\vec \sigma   \otimes I + I \otimes \vec b.\vec \sigma  + \sum\limits_{i,j = 1}^3 {{E_{ij}}{\sigma _i} \otimes {\sigma _j}]},	
\end{equation}
in which $\vec a$ and $\vec b$ are the local Bloch vectors of Alice's and Bob's state respectively, ${E_{ij}} = \mathrm{Tr}[\sigma_i \otimes \sigma_j \rho _{AB}]$ are the elements of correlation matrix $E$, $I= \mathrm{diag}[1, 1, 1, 1]$ denotes an equal mixture of the four Bell states ($\left| {{\psi ^ \pm }} \right\rangle  = ({{\left| {01} \right\rangle  \pm \left| {10} \right\rangle )} \mathord{\left/{\vphantom {{\left| {01} \right\rangle  \pm \left| {10} \right\rangle )}{\sqrt 2 }}} \right.\kern-\nulldelimiterspace} {\sqrt 2 }}$
and $\left| {{\varphi ^ \pm }} \right\rangle  = ({{\left| {00} \right\rangle  \pm \left| {11} \right\rangle )} \mathord{\left/
	{\vphantom {{\left| {00} \right\rangle  \pm \left| {11} \right\rangle )} {\sqrt 2 }}} \right.\kern-\nulldelimiterspace} {\sqrt 2 }}$) and $\vec \sigma   = ({\sigma _x},{\sigma _y},{\sigma _z})$ where ${\sigma _i} (i=x,y,z)$  are Pauli operators.\

In this general model, after restricting the encoding and decoding operators to the class of projective measurements and unitary operators, respectively; the encoding and decoding phases of the protocol reads as follows:\

\textbf{Encoding Phase:} Alice encodes the known signal state into Bob's qubit by performing a projective measurement $\{ \left| { \pm \alpha } \right\rangle \left\langle { \pm \alpha } \right|\}$ on qubit A. This measurement is parameterized by ${\{ {\hat \alpha _m}\} _{m = 1,2}}$, 
where ${\hat \alpha _1} =  + \hat \alpha$, ${\hat \alpha _2} =  - \hat \alpha$ and $\hat \alpha  = ({\alpha _x},{\alpha _y},{\alpha _z})$ is a real unit vector in the Bloch sphere representation. Depending on the output of Alice's measurement, the local Bloch vector of Bob is projected into vector
\begin{equation}\label{bm}
	{\vec b _m} = \frac{{\vec b  + {E^T}{{\hat \alpha }_m}}}{{1 + \vec a .{{\hat \alpha }_m}}},
\end{equation}
	with probability
\begin{equation}\label{pm}
	{P_m} = \frac{1}{2}(1 + \vec a .{\hat \alpha _m}),
\end{equation}
	where $E^T$ is the transpose of the correlation matrix $E$. To activate the quantum channel, Alice should send via a supporting classical channel her measurement result to Bob.\
	
\begin{figure*}[htp]
	\centering
	\includegraphics[scale=0.4]{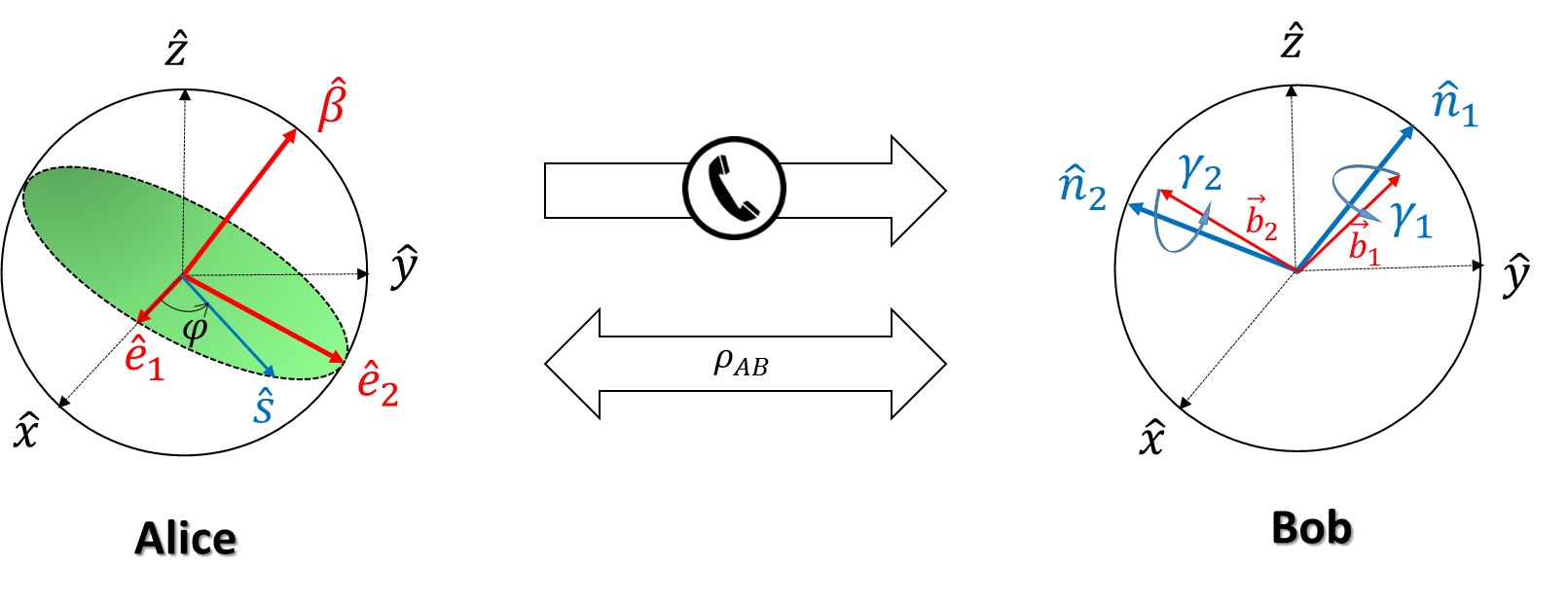}
	\caption{The general scenario of RSP in the Bloch sphere representation. Alice and Bob are equipped with a shared resource state and a supporting classical communication channel. Direction ${\hat \beta }$ is the encoding axis that determines the region chosen by Alice for sending the signal quantum states characterized by the unite vector ${\hat s}$. Depending on the measurement outputs  of Alice, the Bloch vector of Bob's qubit is projected into the vector ${{\vec b}_1}$ or ${{\vec b}_2}$. Alice communicates the output of her measurement ($m=1,2$) and accordingly Bob rotates the projected Bloch vector ${{\vec b}_m}$ about the axis ${{\hat n}_m}$ by the angle ${\gamma _m}$. The unit vectors ${\hat e_1}$ and ${\hat e_2}$ determine the reference frame in the ${\hat s}$ plane. \label{BSR}}
\end{figure*}

\textbf{Decoding Phase:} To decode the signal state, depending on the received classical bit, Bob applies a local unitary transformation, i.e. he applies $U_m$ on his qubit when receives the classical bit $m$. As a result, after receiving the classical bit $m$ and applying the corresponding unitary transformation, Bob decodes the Bloch vector of transmitted state as 
\begin{equation}\label{bmtilda}
	{{\tilde {\vec b}}_m} = {R_m}{{\vec b}_m},
\end{equation}
where $R_m(\hat n_m,\gamma _m)$ is a unique rotation satisfying $U_m(\vec v.\vec \sigma )U_m^\dag  = (R_m\vec v).\vec \sigma$, that rotates any arbitrary Bloch vector $\vec v$ by angle $\gamma_m$ about the direction $\hat n_m$. The explicit form of $R_m$ is given in Appendix \ref{rotationformapp} and the whole scenario is depicted in Fig. \ref{BSR}.
\

After classical communication of Alice's measurement outcome and corresponding decoding operation performed by Bob, by using Eq.(\ref{bm}), Eq.(\ref{pm}) and Eq.(\ref{bmtilda}), the averaged Bloch vector of transmitted state is obtained as 
\begin{equation}\label{r}
	\resizebox{.87\hsize}{!}{$\vec r = \sum\limits_{m = 1,2} {{P_m}{{\tilde {\vec b}}_m} = \frac{1}{2}[({R_1} + {R_2})\vec b + ({R_1} - {R_2}){E^T}\hat \alpha ]}$}.
\end{equation}
The TE of the protocol can be quantified using the standard linear fidelity as follows
\begin{equation}\label{GLF}
	F = \left\langle {{\psi _s}} \right|{\rho ^{(\vec r)}}\left| {{\psi _s}} \right\rangle  = \frac{1}{2}\left( {1 + \vec r.\hat s} \right),
\end{equation}
where ${\rho ^{(\vec r)}}$ is the average prepared state in Bob's site.
Fidelity $F$ in general is a function of input signal state, encoding-decoding and resource state parameters, i.e.
\begin{equation}\label{GLFP}
F = F(\hat s;\underbrace {\hat \alpha ,\hat \beta }_{\mathrm{Enc}};\underbrace {{R_1}\left( {{{\hat n}_1},{\gamma _1}} \right),{R_2}\left( {{{\hat n}_2},{\gamma _2}} \right)}_{\mathrm{Dec}};\underbrace {\vec a,\vec b,E}_{{\mathop{\mathrm{Resource}}\nolimits}}).
\end{equation}
To explore the actual relation between the performance of the protocol and the resource state parameters, all encoding-decoding parameters and the input state in Eq.(\ref{GLFP}) should be traced out in a suitable and meaningful manner. The resulting relation then provides a correspondence between the performance of RSP and the resource state used to perform the task. In the next section we will have an overview of the previous approaches that quantify the performance of the protocol in terms of the resource parameters.

%%%%%%%%%%%%%%%%%%%%%%%%%%%%%%%%%%%%%%%%%%%%%%%%%%%%%%%%

\subsection{Efficiency of the protocol - An overview}\label{Review}
In the first approach, to quantify the performance of RSP, a quadratic fidelity, known as payoff, was defined as $P = {(\vec r.\hat s)^2}$ that after applying some reasonable procedures relates the TE of the protocol to the geometric discord of resource state \cite{dakic2012quantum}. In this approach, for every ${\rho _{AB}}$, the decoding strategy of the protocol is chosen as same as the strategy employed in the standard RSP protocol. In the standard protocol, Alice and Bob agree on a pre-shared direction, characterized by the vector $\hat \beta $. Alice chooses her input states from $\hat s \in \mathrm{GC}(\hat \beta )$ and Bob's decoding operators are ${R_1} = {R_1}(\hat \beta ,0)$ and ${R_2} = {R_2}(\hat \beta ,\pi )$ that correspond to the rotation of Bloch vectors in Bob's coordinate system about the pre-shared $\hat \beta $ direction by $0$ and $\pi$ radian, respectively. After fixing this decoding strategy, for every $\hat s \in \mathrm{GC}(\hat \beta)$ Alice chooses a projective measurement that maximizes the payoff. This maximized payoff is given by

\begin{equation}\label{pmax}
{P_{\mathrm{max}}} = {\left( {E\hat s} \right)^2}.
\end{equation}
In the next step, the maximized payoff is averaged over all $\hat s \in \mathrm{GC}(\hat \beta )$, which gives
\begin{equation}\label{payoff1}
{\cal P}( {\hat \beta } ) = \mathop {{\mathrm{avg}}}\limits_{\hat s{\mathrm{\;}} \in \mathrm{GC}\left( {\hat \beta } \right)}  {{P_{\mathrm{max}}}} = \frac{1}{2}\left(\mathrm{Tr}[ {{E^T}E}] - {{( {E\hat \beta } )}^2} \right),
\end{equation}
where ${\cal P}$ is the resulting TE for the great circle corresponding to the normal vector ${\hat \beta }$. Finally, ${\cal P}( {\hat \beta } )$ is minimized over the direction ${\hat \beta }$ and the resulting minimum payoff is given by
\begin{equation}\label{minpay1}
{{\cal P}_{\min}} = \mathop {\min }\limits_{\hat \beta } {\cal P}( {\hat \beta } ) = {d},
\end{equation}
with
\begin{equation}\label{D1}
{d} = \frac{1}{2}\left( {( {\lambda _1^2 + \lambda _2^2 + \lambda _3^2} ) - \max ( {\lambda _1^2,\lambda _2^2,\lambda _3^2} )} \right),
\end{equation}
where $\lambda _1^2$, $\lambda _2^2$ and $\lambda _3^2$ are eigenvalues of the matrix ${E^T}E$.
Surprisingly, for a broad class of states this expression is equal to the geometric discord of ${\rho _{AB}}$. This result showed that in the worst-case scenario, the TE of the protocol is not less than quantum discord, i.e. discord acts as optimal resource for the RSP protocol.\

However, as remarked in \cite{kanjilal2018remote}, if a zero discord state is used as a resource for RSP, according to the mentioned approach the most one can claim is the existence of at least one great circle, say $\hat \beta_0$, for which RSP cannot be implemented for all $\hat s \in \mathrm{GC}(\hat \beta_0)$. Nevertheless, zero-discord states may still be of use as a resource for
an effective RSP corresponding to the other great circles. Actually, since quantum discord is equal to the minimum of the average payoff, therefore, discord cannot quantify the effectiveness of the resource state in the RSP protocol for any given $\hat \beta $. 
To eliminate this drawback, reference \cite{kanjilal2018remote} suggested that instead of minimizing the averaged payoff over $\hat \beta$ that corresponds to the worst-case scenario, ${\cal P}( {\hat \beta })$ should be averaged over all the great circles characterized by $\hat \beta$. Therefore, one obtains
\begin{equation}\label{Q}
{\cal Q} = \frac{{\smallint d\vec \beta {\cal P}( {\vec \beta } )}}{{\smallint d\vec \beta }} = \frac{1}{3}\left( \lambda _1^2 + \lambda _2^2 + \lambda _3^2 \right).
\end{equation}
Accordingly, if $\cal Q$ vanishes, then ${\cal P}( {\hat \beta } ) = 0$ for all $\hat \beta$ and no signal state can be remotely prepared closer to
the signal state other than the maximally mixed state. Therefore, nonvanishing of the quantity $\cal Q$ can be considered as a necessary condition for the effectiveness of RSP protocol for a given resource state.\

Finally, by considering the recently introduced measures of SCMUB for Bell diagonal states ($\vec a = \vec b =0$), the following analytical relationship between $C_3$ and $\cal Q$ was obtained
\begin{equation}\label{C3andQ}
{C_3}\left( {{\rho _{AB}}} \right) = 1 - h\left( {\frac{{1 + \sqrt {\cal Q} }}{2}} \right),
\end{equation}
where $C_3$ is simultaneous correlations in three mutually unbiased bases introduced in \cite{guo2014quantum} and $h$ is the Shannon function. Interestingly, $C_3$ is a monotonically increasing function of $\cal Q$.
Notice that the quantity $\cal Q$ does not vanish for zero-discord Bell-diagonal states, which are of the form $E = \mathrm{diag}[{\lambda},{0},{0}]$, i.e. zero-discord states are useful as a resource for implementing RSP for a class of signal states pertaining to at least one choice of  $\mathrm{GC}(\hat \beta)$.\

The approaches mentioned in this section use the payoff as quantifier of the TE in RSP protocol. However, there is a need to justify the use of payoff as a  figure of merit to quantify the performance of the protocol. This issue is the subject of the following subsection.

%%%%%%%%%%%%%%%%%%%%%%%%%%%%%%%%%%%%%%%%%%%%%%%%%%%%%%%%

\subsection{Linear fidelity versus quadratic fidelity}\label{LFvsQF}
The payoff function, defined as $P = {(\vec r.\hat s)^2}$, has been used in the previous works to quantify the performance of RSP. However, neither of these works have discussed about the origin of this definition and justification of using payoff as quantifier of the TE in RSP. Such a discussion seems necessary, especially if one considers two states corresponding to the Bloch vectors $+ {{\vec r}_0}$ and $- {{\vec r}_0}$ (with $\vec r.\hat s \ge 0$), respectively. In terms of the linear fidelity, that evidently is a correct figure of merit to quantify the TE of RSP, the states ${\rho ^{( + {{\vec r}_0})}}$ and ${\rho ^{( - {{\vec r}_0})}}$ have different fidelities with respect to a reference signal state, say $\left| {{\psi _{{{\hat s}_0}}}} \right\rangle$; i.e. $F\left( { \pm {{\vec r}_0},{{\hat s}_0}} \right) = \frac{{1 \pm {{\vec r}_0}.{{\hat s}_0}}}{2}$. However, in terms of payoff, clearly we have $P\left( { - {{\vec r}_0},{{\hat s}_0}} \right) = P\left( { + {{\vec r}_0},{{\hat s}_0}} \right)$. This means that payoff is unable to distinguish between orthogonal states ${\rho ^{( - {{\vec r}_0})}}$ and ${\rho ^{( + {{\vec r}_0})}}$ for all ${{\vec r}_0}$. However, we show that after maximizing the payoff over the encoding strategies, this drawback is eliminated. Let us start with the linear fidelity
\begin{equation}\label{GF}
F = \left\langle {{\psi _s}} \right|{\rho ^{(\vec r)}}\left| {{\psi _s}} \right\rangle  = \frac{1}{2}\left( {1 + \vec r.\hat s} \right),
\end{equation}
where ${\vec r}$ is given by Eq.(\ref{r}) and ${\hat s}$ is the Bloch vector of the signal state. In Appendix \ref{optoverencapp} it is shown that after maximizing the linear fidelity over the encoding parameters, i.e. maximizing over ${\hat \alpha }$, the maximized linear fidelity is given by
\begin{equation}\label{FmaxEnc}
\resizebox{.87\hsize}{!}{${F_{\mathrm{max}}} = \frac{1}{2}\left( {1 + \frac{1}{2}\left[ {\left( {{R_1} + {R_2}} \right)\vec b.\hat s + \left|\left| {E\left( {R_1^T - R_2^T} \right)\hat s} \right|\right|} \right]} \right)$},
\end{equation}
where $R_i^T$ is the transpose of the matrix $R_i\,(i = 1,2)$ and $|| {\vec V} ||$ is the Euclidean norm of the vector ${\vec V}$.\

Now, we show that under one of the following conditions, payoff as well as the linear fidelity can successfully quantify the TE of RSP:

\begin{description}\label{C11}
	\item[(C1)] For $\{ {R_1} = {R_1}( {\hat \beta ,0}),{R_2} = {R_2}( {\hat \beta ,\pi } )\}$.
		\item[(C2)] For $\vec b = 0$.
\end{description}
Condition \textbf{(C1)} corresponds to the standard decoding strategy employed in \cite{dakic2012quantum} and condition \textbf{(C2)} corresponds to the resource states with maximally mixed marginal for Bob's qubit.
From Eq.(\ref{FmaxEnc}) it is evident that under one of conditions (\textbf{C1}) or (\textbf{C2}) the first term of Eq.(\ref{FmaxEnc}) vanishes and a maximized encoding strategy leads to ${F_{\mathrm{max}}} \ge \frac{1}{2}$. A comparison between Eq.(\ref{GF}) and Eq.(\ref{FmaxEnc}) shows that, when one of the above conditions is satisfied, just by using a maximized encoding strategy, the Bloch vector of the prepared state always overlaps with the Bloch vector of the signal state by an angel less than $\frac{\pi }{2}$ radian. As a result, one can define the payoff function as
\begin{equation}\label{key}
P \equiv {\left( {F - \frac{1}{2}} \right)^2} = {\left( {\vec r.\hat s} \right)^2},
\end{equation}
with the guarantee that after using an optimal encoding strategy, $\vec r.\hat s \ge 0$ in complete agreement with ${\left( {\vec r.\hat s} \right)^2} \ge 0$, i.e. with an optimized encoding strategy, the state ${\rho ^{( - {{\vec r}_0})}}$ will never be prepared in Bob's site to be mistaken as ${\rho ^{( + {{\vec r}_0})}}$ by the payoff function. Therefore, under conditions \textbf{(C1)} or \textbf{(C2)}, the payoff function and the linear fidelity have similar behavior and both of them can be used to quantify the performance of RSP. More interestingly, as we will see in the next section, under one of the mentioned conditions, the averaged maximum linear fidelity can be expressed in terms of the averaged maximum payoff function and vice versa.\

Before concluding this section, we remark that the argument employed in [14] for invalidating the payoff as figure of merit for the TE of RSP needs more careful consideration. The argument reads as follows \cite{horodecki2014can}:
\begin{quote}
	"Consider a case of the protocol in which Bob, regardless of the Alice message, produces at random a pure state with a Bloch vector belonging to the ${\hat s}$ plane. Employing quadratic fidelity proposed in \cite{dakic2012quantum} one obtains 
	\begin{equation}\label{key}
	\begin{split}
	& {\cal P} = \mathop {\min }\limits_{\hat \beta } {\left( {\hat r.\hat s} \right)^2} = {\left( {\hat r.\hat s} \right)^2}= \smallint d\hat s{\left( {\hat r.\hat s} \right)^2} \\
	&\ \ \ =\frac{1}{{2\pi }}\mathop \smallint \nolimits_0^{2\pi } d\varphi {\cos ^2}\left( \varphi  \right) = \frac{1}{2},
	\end{split}
	\end{equation}
	where the invariance of the measure on the unit
	circle is used. Because this fidelity is higher than those considered in \cite{dakic2012quantum} ( $1/9$ and $1/25$ for separable and entangled state considered in \cite{dakic2012quantum} respectively), it may seem that the random protocol is better choice than more sophisticated strategies. However, this is only because of misleading choice of protocol fidelity."
\end{quote}
However, it is necessary to note that according to the argument presented in the initial paragraphs of this section, only under the optimized encoding strategy the payoff can be regarded as a valid figure of merit to quantify the TE of RSP. Obviously the guessing protocol introduced in the above argument doesn't meet this condition. As a result, when averaging the TE over the unit vector ${\hat s}$, the terms that have negative contributions to the linear fidelity, contribute as positive terms in the payoff that leads to misleading results such as the result obtained in the guessing protocol.\

Therefore, to some extent  the intuition behind the paper \cite{dakic2012quantum} is correct and the challenging result of \cite{dakic2012quantum} cannot be due to the employment of payoff as was claimed by \cite{horodecki2014can}. However, this approach employs a non-optimal decoding strategy. In the next section, the TE of RSP in the fully optimized scenario is investigated.

%%%%%%%%%%%%%%%%%%%%%%%%%%%%%%%%%%%%%%%%%%%%%%%%%%%%%%%%

\subsection{Efficiency of the protocol using the optimized encoding-decoding strategy}\label{fullyoptsection}

In this section we restrict the encoding and decoding maps to the classs of projective measurements and unitary operators, respectively and evaluate the TE of RSP in terms of both quadratic and linear fidelities in the fully optimized scenario (see Fig. \ref{Model}). It is important to note that the optimization over the decoding strategies differs form the optimization over the encoding strategies in the sense that Alice knows the state she intends to transmit and consequently $\forall \hat s \in \mathrm{GC}(\hat \beta )$, she can choose the optimal measurement. However, Bob doesn't know the signal states and for any given $\hat \beta$, the most he can do, is to fix a decoding strategy that maximizes the average fidelity, i.e. for every $\hat \beta$, the TE in terms of the linear fidelity should be defined as follows
\begin{equation}\label{MaxDecF}
{\cal F}( {\hat \beta } ) = \mathop {\max }\limits_{\mathrm{Dec}} {\left\langle {\mathop {\max }\limits_{\mathrm{Enc}} F} \right\rangle _{\hat s \in \mathrm{GC}(\hat \beta )}},
\end{equation}
where ${\left\langle  \bullet  \right\rangle _{_{\hat s \in \mathrm{GC}(\hat \beta )}}}$ stands for averaging the argument over $\hat s \in \mathrm{GC}(\hat \beta )$. The TE in terms of payoff is defined in a similar manner, i.e.
\begin{equation}\label{MaxDecP}
{\cal P}( {\hat \beta } ) = \mathop {\max }\limits_{\mathrm{Dec}} {\left\langle {\mathop {\max }\limits_{\mathrm{Enc}} P} \right\rangle _{\hat s \in \mathrm{GC}(\hat \beta )}}.
\end{equation}
To have a comprehensive approach in recognizing the role of resource state in the TE of RSP it is constructive to define the following expressions,
\begin{equation}\label{pmaxminavgsplit}
\begin{aligned}
	{{\cal F}_{\mathrm{min}}} &= \mathop{\min}\limits_{\hat \beta}{\cal F}({\hat \beta}),\\
	{{\cal F}_{\mathrm{avg}}} &= \mathop{\mathrm{avg}}\limits_{\hat\beta}{\cal F}( {\hat \beta}),\\
	{{\cal F}_{\mathrm{max}}} &= \mathop{\max}\limits_{\hat\beta}{\cal F}({\hat \beta}).
\end{aligned}
\end{equation}

Similar expressions are also defined for payoff. The above expressions also provide a framework for comparing the previous works with the current scenario. 

\begin{figure*}[htp]
	\centering
	\includegraphics[scale=0.8]{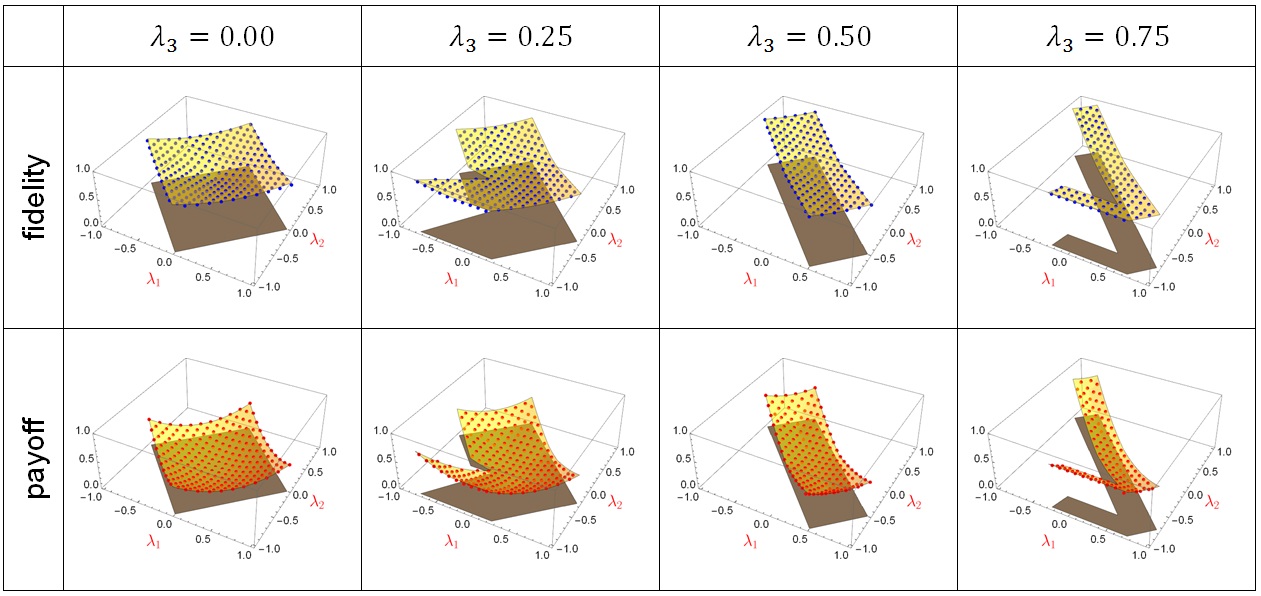}
	\caption{The TE of RSP in terms of the fidelity (linear fidelity) and payoff (quadratic fidelity)  in the fully optimized scenario for Bell-diagonal states. The semi-trasparent surfaces represent the fidelity ${\cal F}( {\hat \beta })$ and payoff ${\cal P}( {\hat \beta })$ given by Eq.(\ref{Fresult}) and Eq.(\ref{Presult}), respectively and the solid balls represent the corresponding numerical results. The numerical results are illustrated for all values of ${\lambda _1}$ and ${\lambda _2}$ belonging to the physical regions (the shaded areas) and four different values of ${\lambda _3}$, i.e. ${\lambda _3} = 0.00,\;0.25,\;0.50,\;0.75$. \label{MainPlot}}
\end{figure*}

The analytical solution to evaluate the above expressions seems impractical at least by formal mathematical analysis methods. Ingenious methods may be later considered by mathematically oriented experts of the field. However, a fully numerical method is also impractical due to the very long execution time of the algorithms. Here, a combined approach is employed, i.e. the optimization over the encoding operators for the linear and quadratic fidelities and also averaging over the input signal states in the quadratic fidelity are performed analytically. Finally, the maximization over the decoding operators and the optimization over the direction ${\hat \beta }$ are performed numerically. All numerical analysis are performed using the standard optimization tools of \textit{Mathematica}. See Appendix for details of the analytical and numerical methods.
The problem is investigated for general resource states. The results show that in the optimized encoding-decoding scenario, the TE of RSP for all  states ${\rho _{AB}}(\vec a,\vec b,E),$ in terms of the linear fidelity is independent of the direction ${\hat \beta }$ and is expressed in terms of the resource state parameters as follows
\begin{equation}\label{Fresult}
\!\!\!{\cal F}( {\hat \beta } ) = \frac{1}{2}\left( {1 + \sqrt {{{\cal D}}} } \right),\ \ \ \ \ \ \ \forall \;\hat \beta  \in S2 
\end{equation}
where $S2$ refers to the surface of the Bloch sphere and
\begin{equation}\label{D2}
{{\cal D}} = \frac{1}{2}\left( {\lambda _1^2 + \lambda _2^2 + \lambda _3^2 - {\mathrm{min}}\left( {\lambda _1^2,\lambda _2^2,\lambda _3^2} \right)} \right),
\end{equation}
where $\lambda _1^2$, $\lambda _2^2$ and $\lambda _3^2$ are eigenvalues of the matrix ${E^T}E$. Fig. \ref{MainPlot} illustrates the numerical results for Bell-diagonal states corresponding to all values of ${\lambda _1}$ and ${\lambda _2}$ belonging to the physical regions and four different values of ${\lambda _3}$, i.e. ${\lambda _3} = 0.00,\;0.25,\;0.50,\;0.75$. The physical regions are obtained from the positivity constraint of the density matrix ${\rho _{AB}}$. Appendix \ref{physicalregionap} gives these physical regions in terms of four inequalities. Note that according to Eq.\eqref{FmaxEnc}, the TE is independent of the local Bloch vector of Alice's state. Also, the TE is independent of the local Bloch vector of Bob's state, because after averaging the maximized linear fidelity over the unit vector $\hat s \in \mathrm{GC}( {\hat \beta })$, the first term in right-hand side of Eq.\eqref{FmaxEnc} vanishes. Hence, the results of Fig. \ref{MainPlot}, are valid for  all states $ {\rho _{AB}}(\vec a,\vec b,E)$. \\

To investigate the TE of RSP in terms of payoff, one should be careful that according to the results of Sec.\ref{LFvsQF}, in the fully optimized scenario, only for states $ {\rho _{AB}}(\vec a,\vec b,E)$ with $\vec b = 0$ payoff is a valid figure of merit to quantify the performance of the protocol. To clarify the justification of this assumption note that, one always can assume that the receiver of the signal state (say Bob) be the distributor of the resource state. Therefore, Bob that starts from a maximally entangled state has control over his qubit with $\vec b = 0$ and the Bloch vector of other part, i.e. $\vec a$, that is transmitted and affected by the channel, according to Eq.(\ref{FmaxEnc}), doesn't contribute to the TE of the protocol. Therefore, in these demanding scenarios, payoff as well as the linear fidelity can successfully quantify the TE of the protocol.\\
The numerical results (see Fig. \ref{MainPlot}) show that in the optimized encoding-decoding scenario, the TE of RSP for all  states ${\rho _{AB}}(\vec a,\vec b,E)$ with $\vec b = 0$ in terms of payoff is independent of the direction ${\hat \beta }$ and is expressed in terms of the resource state parameters as follows
\begin{equation}\label{Presult}
\!{\cal P}( {\hat \beta } ) = {{\cal D}}\ \ \ \ \ \ \ \ \ \ \ \ \ \ \ \ \ \ \ \ \ \;\forall \;\hat \beta  \in S2,
\end{equation}
where, $\cal D$ is given by Eq.(\ref{D2}). From Eq.(\ref{Fresult}) and Eq.(\ref{Presult}) it is evident that the relation between the linear and quadratic fidelities for states ${\rho _{AB}}(\vec a,\vec b,E)$ with $\vec b = 0$ is obtained as follows
\begin{equation}\label{fp}
{\cal F} = \frac{1}{2}\left( {1 + \sqrt {\cal P} } \right).
\end{equation}
This relation is also valid for standard decoding strategy, without requiring $\vec b = 0$.

It is worthy to note that, since the TE in terms of the linear and quadratic fidelities are independent of the direction $\hat \beta$, we will have ${{\cal F}_{\mathrm{min}}} = {{\cal F}_{\mathrm{max}}} = {{\cal F}_{\mathrm{avg}}}$ for the linear fidelity and ${{\cal P}_{\mathrm{min}}} = {{\cal P}_{\mathrm{max}}} = {{\cal P}_{\mathrm{avg}}}$ for the quadratic fidelity. This interesting result shows that by an optimized encoding-decoding strategy, all great circles exploit the resource state equally. Also, from Eq.(\ref{Fresult}) or Eq.(\ref{Presult}) it is evident that any nonproduct state can contribute to the process of RSP, i.e. correlations more than discord serve as a resource for RSP.\

Before concluding this section, let us have a comparison between the result obtained in approach \cite{dakic2012quantum} and result of the present approach. Denoting the obtained fidelity considered in approach \cite{dakic2012quantum}  by ${\cal F}_{\mathrm{min}}^{\left( 1 \right)}$, and also the minimum fidelity obtained in the fully optimized scenario by ${\cal F}_{\mathrm{min}}^{\left( 3 \right)}$; one can check the following inequality,
\begin{equation}\label{compare1}
{\cal F}_{\min}^{\left( 1 \right)} \le {\cal F}_{\min}^{\left( 3 \right)}.
\end{equation}

The strict inequality can be easily understood, as in the new approach a full optimization is carried out. However, the equality are not obvious and need more consideration that is investigated in the next section. 

%%%%%%%%%%%%%%%%%%%%%%%%%%%%%%%%%%%%%%%%%%%%%%%%%%%%%%%%

\subsection{Optimized decoding under special conditions}
Considering Eq.(\ref{fp}), the comparison between Eq.(\ref{minpay1}) and Eq.(\ref{Fresult}) shows that for states with isotropic correlation matrix, i.e. for states with $E =  \pm \lambda I$, the TE of RSP resulting from standard and optimal decoding strategies coincide. Therefore, for isotropic states, $\forall \;\hat \beta  \in S2$ the decoding considered in \cite{dakic2012quantum} and \cite{kanjilal2018remote}, i.e. the decoding strategy used in the standard RSP protocol, is an optimal decoding strategy. Besides this, the comparison between Eq.(\ref{D1}) and Eq.(\ref{D2}) shows that for isotropic states when standard encoding-decoding strategy is employed, quantum
discord truly is optimal resource of RSP.

\begin{figure}[htp]
	\centering
	\includegraphics[scale=0.35]{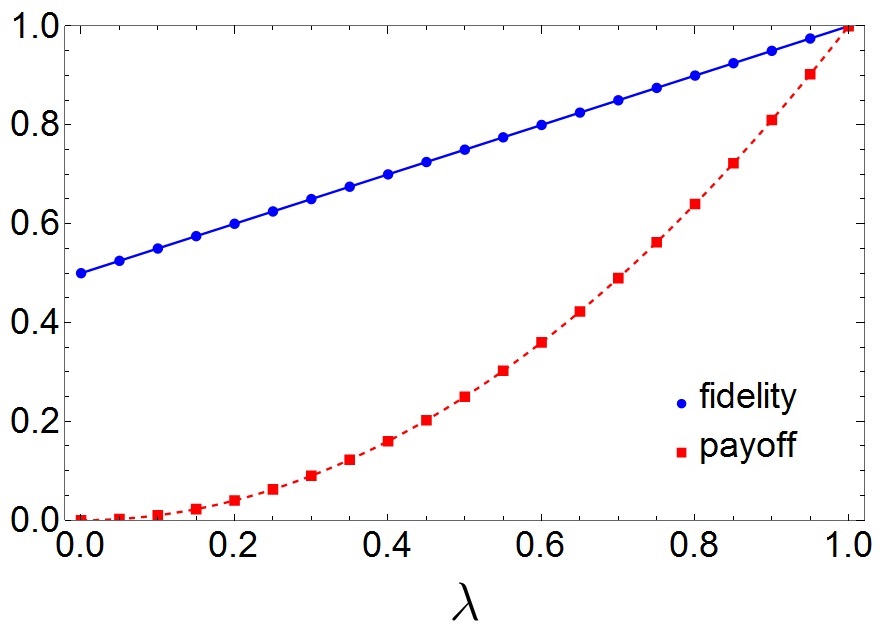}
	\caption{The TE of RSP in terms of the linear and quadratic fidelities for Werner states. The solid line and dotted curve represent the payoff ${\cal P}( {\hat \beta } )$ and fidelity ${\cal F}( {\hat \beta } )$ given by Eq.(\ref{Presult}) (or equivalently by Eq.(\ref{minpay1})) and Eq.(\ref{Fresult}), respectively. The solid squares and solid balls represent the corresponding numerical results.\label{Wernerpic}}
\end{figure}

One important class of isotropic states are Werner states with
\begin{equation}\label{WernerStates}
 \rho _{AB}^{\left( W \right)} = \lambda {\left| {{\psi ^ - }} \right\rangle }\left\langle {{\psi ^ - }} \right| + \frac{{\left( {1 - \lambda } \right)}}{4}I.
\end{equation} 
 Werner states are a family of one parameter states with $\vec a = \vec b = 0,\;E =  - \lambda I$. Hence, for Werner states it is instructive to consider behavior of the TE of RSP in terms of resource state parameters over the full range of $0 \le \lambda  \le 1$. Fig. \ref{Wernerpic} represents the TE of RSP in terms of both linear and quadratic fidelities for this family of states.
\\
It is also interesting to note that for a broad class of states, the decoding strategy employed in the standard RSP protocol, i.e. $({{\hat n}_1} = {{\hat n}_2} = \hat \beta ,{\mathrm{\;}}{\gamma _1} = 0,{\mathrm{\;}}{\gamma _2} = \pi) $, is partially optimum. Actually, in Appendix \ref{parameterizationapp} it is shown that under the constraint ${{\hat n}_1} = {{\hat n}_2} = \hat \beta$, the TE of RSP for all states satisfying the condition $\mathrm{Tr}(EE^T) - {(E\hat \beta )^2}\ge {| {\vec b}|^2} - {(\vec b.\hat \beta )^2}$ (including the states considered in \cite{dakic2012quantum} and \cite{kanjilal2018remote})  takes its maximum when $|{\gamma _1} - {\gamma _2}| = \pi$.

Therefore, the approaches considered in \cite{dakic2012quantum} and \cite{kanjilal2018remote} actually correspond to a scenario with fully optimal encoding strategy and partially optimal decoding strategy and even they correspond to fully optimized scenario for isotropic states.\\

Having characterized various aspects of the approach considered in \cite{dakic2012quantum}, here it's time to concentrate on its challenging result. As shown in \cite{horodecki2014can}, for the most general encoding-decoding CPTP maps, there is no chance for a separable state to beat entanglement efficiency as a resource in any RSP protocol. However, the situation can change when restricting the encoding-decoding operators to some special class of CPTP maps. Following \cite{dakic2012quantum} consider two isotropic states: the separable state ${\rho _{AB}}(\vec a = 0,\vec b = 0,E =  - \frac{1}{3}I)$ and entangled state ${{\rho '}_{AB}}(\vec a = \frac{2}{5}\hat z,\vec b = \frac{2}{5}\hat z,E =  - \frac{1}{5}I)$. From Eq.(\ref{Fresult}) and Eq.(\ref{Presult}) it is evident that the entangled state ${{\rho '}_{AB}}$ has worse RSP fidelity than the separable state ${\rho _{AB}}$ in terms of both linear fidelity and payoff, respectively. Therefore, in the considered class of  encoding-decoding operators, even in their optimal form, there are separable states that perform RSP better than some entangled states. 
%%%%%%%%%%%%%%%%%%%%%%%%%%%%%%%%%%%%%%%%%%%%%%%%%%%%%%%%
%%%%%%%%%%%%%%%%%%%%%%%%%%%%%%%%%%%%%%%%%%%%%%%%%%%%%%%%

\section{Discussion}
Communication is nothing but making correlation between sent and received data via communication channels. It is obvious that for efficient transmitting of quantum information, the employment of quantum communication channels is necessary. A correlated shared state supported with a classical communication channel, acts as a quantum channel. Here, the scenario of RSP with a shared correlated quantum state ${\rho _{AB}}$ and one bit of forward
communication was considered. The correlation capability of this quantum channel (i.e. ${\rho _{AB}}$) can be explicitly expressed as
\begin{equation}\label{kay}
\resizebox{.87\hsize}{!}{$\chi := {\rho _{AB}} - {\rho _A} \otimes {\rho _B} =\frac{1}{4} \sum\limits_{i,j} {({E_{ij}} - {a_i}{b_j})} ({\sigma _i} \otimes {\sigma _j})$},
\end{equation}
where ${\rho _{A,B}} = {\mathrm{Tr}_\mathrm{B,A}}(\rho _{AB})$ contains the local information of subsystem $\mathrm A$ (subsystem $\mathrm B$) and $a_i$ ($b_i$) is the $i$-th component of its corresponding Bloch vector. From Eq.(\ref{kay}) it is evident  that the correlation capability of the channel is determined by $E-\vec a \vec b$. The role of $E$ in the TE of RSP becomes apparent when considering Eq.(\ref{bm}) that shows, the correlation matrix $E$ is responsible for transmitting the encoded state characterized by ${{\hat \alpha }_m}$. However, as a result of averaging over the input and output states, the TE of RSP doesn't depend on the local Bloch vectors $\vec b$ and $\vec a$, respectively. This is why the TE of RSP is obtained as a function of parameters of the correlation matrix. However, the main part of the problem is to recognize the optimal resource of the RSP.\

The approach considered in \cite{dakic2012quantum} showed that for every great circle, the geometric discord is a lower bound for the TE of RSP. However, this approach cannot quantify the performance of the protocol for every great circle. Also, this approach cannot quantify the usefulness of resource states with vanishing discord and is unable to compare the usefulness of two resource states for RSP. In order to improve these drawbacks, evaluation of average performance of the protocol was suggested in \cite{kanjilal2018remote}. This idea characterized the average performance of the protocol in terms of correlations beyond discord which showed some of classical correlations (if we take quantum discord as quantum correlations) are also resource for RSP. However, in this approach, averaging is over the direction ${\hat \beta }$ that corresponds to averaging over different arrangements of the protocol. Note that for every arrangement of the protocol, Alice and Bob should agree on the pre-shared direction ${\hat \beta }$ that in turn requires its own resources (see for example \cite{chiribella2004efficient}, \cite{van2005quantifying}). Therefore, this approach is not justified in the perspective of the resource theory. \
  
The present approach shows that when the protocol is equipped with the optimized decoding strategy, correlations more than quantum discord are activated to serve as a resource for RSP. Furthermore, in the fully optimized scenario, the drawbacks of \cite{dakic2012quantum} are automatically eliminated. Actually, from Eq.(\ref{Presult}) it is evident that if ${\cal {D}}=0$, then there are no states that can be remotely prepared closer to the signal state other than the maximally mixed state, which indicates the ineffectiveness of RSP protocol corresponding to any value of ${\hat \beta }$; i.e. ${\cal P}( {\hat \beta } ) = 0,\;\forall \;\hat \beta  \in S2$. On the other hand, nonvanishing of the quantity $\cal {D}$ can be taken as a necessary condition for usefulness of a given resource state for the RSP task corresponding to any great circle. Furthermore, Eq.(\ref{Fresult}) (or Eq.(\ref{Presult})) quantifies the performance of the RSP protocol for every great circle and  provides a means to compare the usefulness of two resource states for performing the RSP task.\

Although in this approach the TE of RSP is characterized in terms of the quantity $\cal {D}$, this remains as an open problem that what kind of quantumness captures this amount of correlation. Usually, for a given quantum correlation measure an operational interpretation is presented. Here the situation is reversed. The optimal TE of RSP task is given by the sum of two largest eigenvalues of the squared correlation matrix of the resource state and the question is that what type of quantum correlations is related to this quantity. 

%%%%%%%%%%%%%%%%%%%%%%%%%%%%%%%%%%%%%%%%%%%%%%%%%%%%%%%%
%%%%%%%%%%%%%%%%%%%%%%%%%%%%%%%%%%%%%%%%%%%%%%%%%%%%%%%%

\section{Summary}
In summary, the transmission efficiency (TE) of remote state preparation (RSP) with a shared quantum state and one bit of classical communication was considered.  Following \cite{dakic2012quantum}, the encoding and decoding operators of the protocol was restricted to the classes of projective measurements and unitary operators, respectively. In this context, the previous approaches to the problem were reviewed, analyzed and improved.
First, it was shown that contrary to the argument given in \cite{horodecki2014can}, the quadratic fidelity as well as the linear fidelity could be a valid figure of merit to quantify the TE of RSP. Hence, the challenging result of \cite{dakic2012quantum}, i.e. certain entangled states can be outperformed by some states without any entanglement, cannot be due to the employment of the quadratic fidelity. Then, the TE of RSP was evaluated in a fully optimized scenario which includes the maximization over the encoding parameters as well as a meaningful maximization over the decoding parameters. In this scenario, the TE in terms of both linear and quadratic fidelities scales with the sum of two largest eigenvalues of the squared correlation matrix of the resource state that is zero only for product states. This result is independent of the input region (the great circle in the Bloch sphere representation) chosen by Alice for sending the signal quantum states. Hence, this approach doesn't have the drawbacks of the approach considered in \cite{dakic2012quantum}, i.e. for any great circle chosen by Alice, it precisely determines the usefulness of a resource state for the RSP task and also it can be used to compare the usefulness of two resource states for RSP task corresponding to any great circle. 
The mentioned drawbacks have been eliminated in another approach \cite{kanjilal2018remote}, where the standard decoding strategy was employed and instead of evaluating the TE in the worst case scenario, the average performance of the protocol was considered. However as discussed, averaging over different great circles corresponds to averaging over different arrangements of the protocol that is not justified in the perspective of the resource theory.\
Furthermore, it was shown that in the considered class of encoding-decoding operators, i.e. projective measurements and unitary operators, even in their optimal form, there are separable states that outperform entangled state in RSP task. Also, it was shown that the standard decoding strategy employed in the previous approaches is partially optimum. In particular it is fully optimum for isotropic resource states.\

\begin{acknowledgments}
	This work was supported by the Center for Quantum Engineering and Photonics Technologies at Sharif University of Technology (SUT) through the Optical Quantum Communication Project. We would also like to give special thanks to the members of the quantum communication group at SUT for useful discussions.
\end{acknowledgments}

%%%%%%%%%%%%%%%%%%%%%%%%%%%%%%%%%%%%%%%%%%%%%%%%%%%%%%%%
%%%%%%%%%%%%%%%%%%%%%%%%%%%%%%%%%%%%%%%%%%%%%%%%%%%%%%%%

\appendix*
\section{(Methods)}
After Local operation of Bob, coordinated with classical communication of Alice, the average remotely prepared state of Bob's qubit is obtained as follows
\begin{equation}\label{meanstate}
\rho _B^{\left( {\vec r} \right)} = \mathop \sum \limits_{m = 1,2} {P_m}{U_m}\left( {\frac{{I + {{\vec b}_m}.\vec \sigma }}{2}} \right)U_m^\dag,
\end{equation}
where ${U_m} = {e^{ - i\frac{{{\gamma _m}}}{2}{{\hat n}_m}.\vec \sigma }};(m = 1,2)$ is the general rotation operator in {${\mathbb{C}}^{2}$}, ${{\vec b}_m}$ and $P_m$ are given by Eq.(\ref{bm}) and Eq.(\ref{pm}),  respectively and $\vec \sigma  = \left( {{\sigma _x},{\sigma _y},{\sigma _z}} \right)$. Using properties of Pauli matrices, standard calculations leads to the following expression for Bob's qubit state
\begin{equation}\label{Bobsqubitstate}
\rho _B^{\left( {\vec r} \right)} = \frac{{I + \vec r.\vec \sigma }}{2},
\end{equation}
with
\begin{equation}\label{Bobsqubitaverager}
\begin{split}
& \vec r = \mathop \sum \limits_m {P_m}(\cos \left( {{\gamma _m}} \right){{\vec b}_m} + \sin \left( {{\gamma _m}} \right){{\hat n}_m} \times {{\vec b}_m}\\
& + \left( {1 - \cos \left( {{\gamma _m}} \right)} \right){{\hat n}_m}{{\hat n}_m}.{{\vec b}_m}).
\end{split}
\end{equation}

%%%%%%%%%%%%%%%%%%%%%%%%%%%%%%%%%%%%%%%%%%%%%%%%%%%%%%%%

\subsection{The explicit form of the rotation operator}\label{rotationformapp}
Eq.(\ref{Bobsqubitaverager}) can be rewritten as follows
\begin{equation}\label{avgrintermsofRm}
\vec r = \mathop \sum \limits_m {P_m}{R_m}{{\vec b}_m},
\end{equation}
where $R_m$ is a $3\times3$ rotation matrix with the following elements
\begin{equation}
\begin{split}
& R_m^{(x,x)} = n_m^{(x)}n_m^{(x)}\left( {1 - \cos \left( {{\gamma _m}} \right)} \right) + {\mathrm{cos}}\left( {{\gamma _m}} \right), \\
& R_m^{(x,y)} = n_m^{(x)}n_m^{(y)}\left( {1 - \cos \left( {{\gamma _m}} \right)} \right) + {\mathrm{sin}}\left( {{\gamma _m}} \right)n_m^{(z)},\\
& R_m^{(x,z)} = n_m^{(x)}n_m^{(z)}\left( {1 - \cos \left( {{\gamma _m}} \right)} \right) + {\mathrm{sin}}\left( {{\gamma _m}} \right)n_m^{(y)},\\
& R_m^{(y,x)} = n_m^{(y)}n_m^{(x)}\left( {1 - \cos \left( {{\gamma _m}} \right)} \right) + {\mathrm{sin}}\left( {{\gamma _m}} \right)n_m^{(z)},\\
& R_m^{(y,y)} = n_m^{(y)}n_m^{(y)}\left( {1 - \cos \left( {{\gamma _m}} \right)} \right) + {\mathrm{cos}}\left( {{\gamma _m}} \right),\\
& R_m^{(y,z)} = n_m^{(y)}n_m^{(z)}\left( {1 - \cos \left( {{\gamma _m}} \right)} \right) - {\mathrm{sin}}\left( {{\gamma _m}} \right)n_m^{(x)},\\
& R_m^{(z,x)} = n_m^{(z)}n_m^{(x)}\left( {1 - \cos \left( {{\gamma _m}} \right)} \right) - {\mathrm{sin}}\left( {{\gamma _m}} \right)n_m^{(y)},\\
& R_m^{(z,y)} = n_m^{(z)}n_m^{(y)}\left( {1 - \cos \left( {{\gamma _m}} \right)} \right) + {\mathrm{sin}}\left( {{\gamma _m}} \right)n_m^{(x)},\\
& R_m^{(z,z)} = n_m^{(z)}n_m^{(z)}\left( {1 - \cos \left( {{\gamma _m}} \right)} \right) + {\mathrm{cos}}\left( {{\gamma _m}} \right),\\
\end{split}\label{Rmelements}
\end{equation}
where $R_m^{(i,j)}$ is the $(i,j)$-th element of the matrix $R_m$ and $n_m^{(k)}$ is the $k$-th element of the unit vector ${{\hat n}_m}$.

%%%%%%%%%%%%%%%%%%%%%%%%%%%%%%%%%%%%%%%%%%%%%%%%%%%%%%%%

\subsection{Optimization over the encoding}\label{optoverencapp}
By substituting Eq.(\ref{bm}) and Eq.(\ref{pm}) into Eq.(\ref{avgrintermsofRm}) one gets
\begin{equation}\label{r.s}
\vec r.\hat s = \frac{1}{2}\left[ {\left( {{R_1} + {R_2}} \right)\vec b.\hat s + \left( {{R_1} - {R_2}} \right){E^T}\hat \alpha .\hat s} \right],
\end{equation}
which equivalently can be written as
\begin{equation}\label{transposedr.s}
\vec r.\hat s = \frac{1}{2}\left[ {\left( {\left( {{R_1} + {R_2}} \right)\vec b.\hat s + \hat \alpha .E\left( {R_1^T - R_2^T} \right)\hat s} \right)} \right],
\end{equation}
where $R_i^T\;(i = 1,2)$ is the transpose of the rotation matrix $R_i$. Now, since $\left|\left| {\hat \alpha } \right|\right| = 1$, $\vec r.\hat s$ is maximized when
 \begin{equation}\label{maxalpha}
 \hat \alpha  = \frac{{E\left( {R_1^T - R_2^T} \right)\hat s}}{{\left|\left| {E\left( {R_1^T - R_2^T} \right)\hat s} \right|\right|}}.
 \end{equation}
 Therefore, the maximum of $\vec r.\hat s$ over ${\hat \alpha }$ leads to the following quantity
\begin{equation}\label{finalencmax}
\resizebox{.85\hsize}{!}{$\mathop {\max }\limits_{\hat \alpha } (\vec r.\hat s) = \frac{1}{2}\left[ {\left( {{R_1} + {R_2}} \right)\vec b.\hat s + \left| \left|{E\left( {R_1^T - R_2^T} \right)\hat s} \right|\right|} \right]$}.
\end{equation}

%%%%%%%%%%%%%%%%%%%%%%%%%%%%%%%%%%%%%%%%%%%%%%%%%%%%%%%%

\subsection{Parameterization of Bloch vector of the signal state}\label{parameterizationapp}
Evaluation of the fidelity requires averaging the quantity given by Eq.(\ref{finalencmax}) over all signal states chosen from the great circle characterized by the unit vector ${\hat \beta }$ in Bloch sphere representation, i.e.
\begin{equation}\label{recallfidelity}
{\cal F}( {\hat \beta }) = \frac{1}{2}(1 + {\langle {\mathop {\max }\limits_{\hat \alpha } (\vec r.\hat s)} \rangle _{\hat s\; \in \mathrm{GC}\left( {\hat \beta } \right)}}).
\end{equation}
For doing so, the unit vector ${\hat s}$ requires to be parameterized so as to characterize the great circle.
Parameterization of the unit vector $\hat s \in \mathrm{GC}( {\hat \beta })$ is performed as follows. Suppose that the unit vectors ${\hat e}_1$ and ${\hat e}_2$ determine the reference frame in the ${\hat s}$ plane. The unit vector ${\hat e}_1$ is determined as intersection of the great circle plane and $xy$ plane. Then, the unit vector ${\hat e}_2$ is chosen such that ${{\hat e}_1}.{{\hat e}_2}=\hat \beta.{{\hat e}_1}=\hat \beta .{{\hat e}_2} = 0,$ and ${{\hat e}_1} \times {{\hat e}_2} = \hat \beta $. Solving all these equations simultaneously leads to following expression for the Bloch vector of the signal state
\begin{equation}
\begin{split}
& {s_x} =  - {\mathrm{cos}}\left( {{\theta _\beta }} \right){\mathrm{cos}}\left( {{\phi _\beta }} \right)sin\left( \phi  \right) - {\mathrm{cos}}\left( \phi  \right){\mathrm{sin}}\left( {{\phi _\beta }} \right),\\ 
& {s_y} = {\mathrm{cos}}\left( \phi  \right){\mathrm{cos}}\left( {{\phi _\beta }} \right) - {\mathrm{cos}}\left( {{\theta _\beta }} \right){\mathrm{sin}}\left( \phi  \right){\mathrm{sin}}\left( {{\phi _\beta }} \right),\\
& {s_z} = \sin \left( {{\theta _\beta }} \right)\sin \left( \phi  \right),\\
\end{split}\label{sparameterscomponent}
\end{equation}
where ${\theta _\beta }$ and ${{\phi _\beta }}$ are components of the vector ${\hat \beta }$ in  spherical coordinate system and $\phi $ is an angel measured from the vector ${{{\mathrm{\hat e}}}_1}$ that determines unit vector ${\hat s}$ in $\mathrm {GC}({\hat \beta })$.
This parameterization is used as well to evaluate the average payoff. 

%%%%%%%%%%%%%%%%%%%%%%%%%%%%%%%%%%%%%%%%%%%%%%%%%%%%%%%%

\subsection{Partial optimization}\label{partialoptapp}
Here, it is shown that for all general resource states, the decoding strategy employed in the standard RSP protocol i.e. ${{\hat n}_1} = {{\hat n}_2} = \hat \beta ,{\mathrm{\;}}{\gamma _1} = 0,{\mathrm{\;}}{\gamma _2} = \pi$, is partially optimum. 
According to Eq.(\ref{finalencmax}) the payoff after maximizing over the encoding can be written as
\begin{equation}\label{key}
{P_{\max}} = \frac{1}{4}{\left[ {\left( {{R_1} + {R_2}} \right)\vec b.\hat s + \left|\left| {E\left( {R_1^T - R_2^T} \right)\hat s} \right|\right|} \right]^2}.
\end{equation}
After averaging $P_{\mathrm{max}}$ over $\hat s(\phi) \in \mathrm{GC}(\hat \beta)$ parameterized in Eq.(\ref{sparameterscomponent}),  straightforward but lengthy calculation leads to the following expression
\begin{equation}\label{optpartialdec}
\begin{aligned}
{{\cal P}_c}( {\hat \beta } ) = \frac{1}{2}\left( {{{| {\vec b} |}^2} - {{( {\vec b.\vec \beta } )}^2}} \right){\cos ^2}(\frac{{{\gamma _1} - {\gamma _2}}}{2}) \\
+ \frac{1}{2}\left( \mathrm{Tr}[ {{E^T}E}] - {{( {E\hat \beta } )}^2} \right){\sin ^2}(\frac{{{\gamma _1} - {\gamma _2}}}{2}),
\end{aligned}
\end{equation}
where ${{\cal P}_c}$ denotes the resulting payoff with decoding constraint ${{\hat n}_1} = {{\hat n}_2} = \hat \beta$.
From Eq.(\ref{optpartialdec}) it is evident that under the condition $ \mathrm{Tr}[E{E^T}] - {(E\hat \beta )^2}$$ \ge {| {\vec b}|^2} - {(\vec b.\hat \beta )^2}$, ${{\cal P}_c}$ takes its maximum when $|{\gamma_2} - {\gamma_1}| = \pi$, and under the condition $\mathrm{Tr}[E{E^T}] - {(E\hat \beta )^2} \le {| {\vec b}|^2} - {(\vec b.\hat \beta )^2}$, it takes the maximum when $ |{\gamma_2} - {\gamma_1}| = 0$.
\subsection{The physical regions of the states}\label{physicalregionap}
The positivity condition of density matrices restricts the eigenvalues of the correlation matrix $E$ to a physical region constrained by the following four inequalities
\begin{equation}\label{key}
\begin{split}
& 1 - {\lambda _1} - {\lambda _2} - {\lambda _3} \ge 0,\ \ 1 - {\lambda _1} + {\lambda _2} + {\lambda _3} \ge 0,\\
& 1 + {\lambda _1} - {\lambda _2} + {\lambda _3} \ge 0,\ \ 1 + {\lambda _1} + {\lambda _2} - {\lambda _3} \ge 0,\\
\end{split}
\end{equation}
\\
which is equivalent to a tetrahedron ${\cal T}$ with vertices ${t_0} = \left( { - 1, - 1, - 1} \right)$, ${t_1} = \left( { - 1,  1, 1} \right)$, ${t_2} = \left( {  1, - 1,  1} \right)$, ${t_3} = \left( {  1, 1, - 1} \right)$ \cite{horodecki1996information}.

\bibliography{revised}

\end{document}